# Kepler-16: A Transiting Circumbinary Planet

Laurance R. Doyle[1], Joshua A. Carter[2], Daniel C. Fabrycky[3], Robert W. Slawson[1], Steve B. Howell[4], Joshua N. Winn[5], Jerome A. Orosz[6], Andrej Prša[7], William F. Welsh[6], Samuel N. Quinn[8], David Latham[8], Guillermo Torres[8], Lars A. Buchhave[9,10], Geoffrey W. Marcy[11], Jonathan J. Fortney[12], Avi Shporer[13,14], Eric B. Ford[15], Jack J. Lissauer[4], Darin Ragozzine[2], Michael Rucker[16], Natalie Batalha[16], Jon M. Jenkins[1], William J. Borucki[4], David Koch[4], Christopher K. Middour[17], Jennifer R. Hall[17], Sean McCauliff[17], Michael N. Fanelli[18], Elisa V. Quintana[1], Matthew J. Holman[8], Douglas A. Caldwell[1], Martin Still[18], Robert P. Stefanik[8], Warren R. Brown[8], Gilbert A. Esquerdo[8], Sumin Tang[8], Gabor Furesz[8,19], John C. Geary[8], Perry Berlind[20], Michael L. Calkins[20], Donald R. Short[21], Jason H. Steffen[22], Dimitar Sasselov[8], Edward W. Dunham[23], William D. Cochran[24], Alan Boss[25], Michael R. Haas[4], Derek Buzasi[26], Debra Fischer[27]

[1]Carl Sagan Center for the Study of Life in the Universe, SETI Institute, 189 Bernardo Avenue, Mountain View, CA 94043, USA, ldoyle@seti.org, rslawson@seti.org, jon.jenkins@nasa.gov, elisa.quintana@nasa.gov, douglas.a.caldwell@nasa.gov

[2]Hubble Fellow, Harvard-Smithsonian Center for Astrophysics, 60 Garden Street, Cambridge, MA 02138, USA, jacarter@cfa.harvard.edu

[3]Hubble Fellow, Department of Astronomy and Astrophysics, University of California, Santa Cruz, CA 95064, USA, fabrycky@ucolick.org

[4]NASA Ames Research Center, Moffett Field, CA 94035, USA, steve.b.howell@nasa.gov, jack.j.lissauer@nasa.gov, william.j.borucki@nasa.gov, d.koch@nasa.gov

[5]Massachusetts Institute of Technology, Physics Department and Kavli Institute for Astrophysics and Space Research, 77 Massachusetts Avenue, Cambridge, MA 02139, USA, jwinn@mit.edu

[6]Astronomy Department, San Diego State University, 5500 Campanile Drive, San Diego, CA 92182-1221, orosz@sciences.sdsu.edu, wfw@sciences.sdsu.edu

[7]Villanova University, Dept. of Astronomy and Astrophysics, 800 E Lancaster Ave, Villanova, PA 19085, USA, andrej.prsa@villanova.edu

[8]Harvard-Smithsonian Center for Astrophysics, 60 Garden Street, Cambridge, MA 02138, USA, dlatham@cfa.harvard.edu, gtorres@cfa.harvard.edu, mholman@cfa.harvard.edu, rstefanik@cfa.harvard.edu, wbrown@cfa.harvard.edu, esquerdo@psi.edu, stang@cfa.harvard.edu, gfuresz@cfa.harvard.edu, jgeary@cfa.harvard.edu, dsasselov@cfa.harvard.edu

[9]Niels Bohr Institute, Copenhagen University, DK-2100 Copenhagen, Denmark, buchhave@astro.ku.dk

[10]Centre for Star and Planet Formation, Natural History Museum of Denmark, University of Copenhagen, DK-1350 Copenhagen, Denmark

[11]Astronomy Department, University of California, Berkeley, CA, 94720, USA, gmarcy@berkeley.edu

[12]Department of Astronomy and Astrophysics, University of California, Santa Cruz, Santa Cruz, CA 95064, USA, jfortney@ucolick.org

[13]Las Cumbres Observatory Global Telescope Network, 6740 Cortona Drive, Suite 102, Santa Barbara, CA 93117, USA,

[14]Department of Physics, Broida Hall, University of California, Santa Barbara, CA 93106, USA, shporer@lcogt.net

[15]211 Bryant Space Science Center, Gainesville, FL 32611-2055, USA, eford@astro.ufl.edu

[16]Physics Department, San Jose State University, San Jose, CA, 95192, USA, natalie.m.batalha@nasa.gov, mrucker967@gmail.com

[17]Orbital Sciences Corporation/NASA Ames Research Center, Moffett Field, CA 94035, USA, christopher.k.middour@nasa.gov, jennifer.hall@nasa.gov, sean.d.mccauliff@nasa.gov

[18]Bay Area Environmental Research Inst./NASA Ames Research Center, Moffett Field, CA 94035, USA michael.n.fanelli@nasa.gov, martin.still@nasa.gov

[19]Konkoly Observatory, Konkoly ut 15-17, Budapest, H-1121, Hungary

[20]Fred Lawrence Whipple Observatory, Smithsonian Astrophysical Observatory, Amado, AZ 85645, USA, pberlind@cfa.harvard.edu, mcalkins@cfa.harvard.edu

[21]Mathematics Department, San Diego State University, 5500 Campanile Drive, San Diego, CA USA 92182, dshort@sciences.sdsu.edu

[22]Fermilab Center for Particle Astrophysics, P.O. Box 500, Batavia IL 60510, USA jsteffen@fnal.gov

[23]Lowell Observatory, Flagstaff, AZ, 86001, USA, Dunham@lowell.edu
[24]McDonald Observatory, University of Texas at Austin, Austin, TX, 78712, USA, wdc@astro.as.utexas.edu
[25]Carnegie Institute of Washington, Washington, DC 20015 USA, boss@dtm.ciw.edu
[26]Eureka Scientific, Inc., 2452 Delmer Street Suite 100, Oakland, CA 94602, USA, dbuzasi@gmail.com
[27]Department of Astronomy, Yale University, New Haven, CT 06511 USA, debra.fischer@gmail.com

**Abstract**

*We report the detection of a planet whose orbit surrounds a pair of low-mass stars. Data from the Kepler spacecraft reveal transits of the planet across both stars, in addition to the mutual eclipses of the stars, giving precise constraints on the absolute dimensions of all three bodies. The planet is comparable to Saturn in mass and size, and is on a nearly circular 229-day orbit around its two parent stars. The eclipsing stars are 20% and 69% as massive as the sun, and have an eccentric 41-day orbit. The motions of all three bodies are confined to within 0.5° of a single plane, suggesting that the planet formed within a circumbinary disk.*

A planet with two suns is a familiar concept from science fiction. However, the evidence for the existence of circumbinary planets—those that orbit around both members of a stellar binary—has been limited. A few good cases have been made for circumbinary planets based upon timing of stellar eclipses (see, e.g., refs. *1-3*), but in no previous case have astronomers obtained direct evidence of a circumbinary planet by observing a planetary transit (a miniature eclipse as the planet passes directly in front of a star). Detection of a transit greatly enhances confidence in the reality of the planet, and provides unusually precise knowledge of its mass, radius, and orbital parameters (*4*).

Here we present the detection of a transiting circumbinary planet around a binary star system based on photometric data from the NASA *Kepler* spacecraft. *Kepler* is a 0.95m space telescope that monitors the optical brightness of about 155,000 stars within a field encompassing 105 square degrees in the constellations Cygnus and Lyra (*5-8*).

Star number 12644769 from the *Kepler* Input Catalog was identified as an eclipsing binary with a 41-day period, from the detection of its mutual eclipses (*9*). Eclipses occur because the orbital plane of the stars is oriented nearly edge-on as viewed from Earth. During primary eclipses the larger star, denoted "A", is partially eclipsed by the smaller star "B", and the system flux declines by about 13%. During secondary eclipses B is completely occulted by A, and the resulting drop in flux is only about 1.6% because B is relatively small and has a lower surface brightness (Figure 1).

This target drew further attention when three additional drops in brightness were detected outside of the primary and secondary eclipses, separated by intervals of 230.3 and 221.5 days (*10*). These tertiary eclipses could not be attributed to the stars alone, and indicated the presence of a third body. The differing intervals between the tertiary eclipses are simply explained if the third body is in a circumbinary orbit, because stars A and B would be in different positions in their mutual orbit each time the third body moved in front of them (*11, 12*). In contrast, there would be no ready explanation for the shifting times of the tertiary eclipses if they were produced by a background star system or some other unrelated event.

During tertiary eclipses the total light declines by 1.7%. Because this is larger than the 1.6% decline during secondary eclipses (when star B is completely concealed), the tertiary eclipses had to be transits of the third body across star A. This interpretation was supported by the subsequent detection of weaker 0.1% quaternary eclipses, which were consistent with the passage of the third body across star B. The observed time of this quaternary eclipse was used to predict two other times of quaternary eclipses that should have been present in the data, and these two events were subsequently detected (Figure 1).

Because the third body covers only 1.7% of the area of star A, which was determined to be smaller than the Sun based on its broad band colors (*10*), the circumbinary body was suspected to be either a planet, or a third star with grazing eclipses. Decisive evidence that it is a planet came from investigation of the timing of the stellar eclipses. The primary and secondary eclipse times were found to depart from strict periodicity by deviations of order one minute. A third body causes timing variations in two ways. Firstly there is a light travel-time effect: the third body induces a periodic motion of the center of mass of the stellar binary, causing periodic variations in the time required for the eclipse signals to reach the Earth (*13, 14*). Secondly there is a dynamical effect: the gravitational attraction of each star to the third body varies with time due to the changing positions of all three bodies, causing perturbations in the stars' orbital parameters and therefore in the eclipse times (*15, 16*). Both effects depend on the mass of the third body. Therefore we could constrain the mass of the third body by fitting the eclipse data with a numerical model of three-body gravitational interactions. This model, described below in detail, showed that the third body must be less massive than Jupiter.

Hence, based on the depth of the tertiary eclipses, and on the magnitude of the eclipse timing variations, the third body was shown to be a transiting circumbinary planet.

The model was based on the premise that the three bodies move under the influence of mutual Newtonian gravitational forces. For this purpose we modified the computer code that was used to model the triple star system KOI-126 (*17*, SOM). The leading-order relativistic correction to the force law was included, although it proved to be unimportant. The bodies' positions were calculated with a Bulirsch-Stoer algorithm and corrected for the finite propagation speed of light across the system before comparing to the data. The loss of light due to eclipses was calculated by assuming the disks of stars A and B to be circular, with a quadratic law describing the decline in intensity toward the limb (*18*). We also allowed for an additional time-independent source of light to account for any possible background stars within the *Kepler* photometric aperture. In practice this parameter was found to be consistent with zero, and bounded to be less than 1.3% of the total light of the system (*19*).

We fitted all of the photometric data within 6 hours of any eclipse or transit. Before fitting, a linear trend was removed from each segment, to correct for the slow starspot-induced variations evident in Figure 1. A successful model had to be compatible with the timings, durations, and depths of the primary and secondary stellar eclipses, as well as the transits of the planet across both stars. The model also had to account for the slight departures from strict periodicity of the stellar eclipses. Furthermore, to pin down the stellar masses and provide an absolute distance scale, we undertook spectroscopic observations to track the radial velocity variations of star A (Figure 2, top panel).

The model parameters were adjusted to fit the photometric and radial-velocity data (Table 1). Figures 1 and 2 show the very good match that was achieved between the model and the data. Uncertainties in the parameters were determined with a Differential Evolution Markov Chain Monte Carlo simulation (*21,* SOM).

Due to the presence of uniquely three-body effects (namely, the shifts in eclipse times and transit durations), the masses, radii, and orbital distances of this system are well determined in absolute units, and not just in relative units. The eclipse timing variations are dominated by the effects of dynamical perturbations, with light-time variations contributing only at the level of one second. The third body's dimensions are well within the planetary regime, with a mass of $0.333 \pm 0.016$ and a radius of $0.7538 \pm 0.0025$ those of Jupiter. Following the convention of Ref. 22, we can denote the third body Kepler-16 (AB)-b, or simply "b" when there is no ambiguity.

Considering its bulk properties, the planet is reminiscent of Saturn but with a higher mean density ($0.964$ g cm$^{-3}$, compared to the Saturnian density of $0.687$ g cm$^{-3}$). This suggests a greater degree of enrichment by heavy elements. With a mass and radius one can begin to model a planet's interior structure, which will depend on age because planets cool and contract with time. Usually the stellar age is used as a proxy for the planetary age, but in this case the stellar age is not unambiguous. The primary star is a slow rotator (with a period of about 35.1 days, judging from the out-of-eclipse variations), usually indicative of old age. In contrast, the level of starspot activity and chromospheric emission (Mt. Wilson *S* value = 1.10) are indicative of youth. The spectroscopic determination of star A's heavy-element fraction ([m/H] = $-0.3 \pm 0.2$) is also relatively uncertain, making it more difficult to estimate the age with theoretical evolutionary models. Nevertheless, for any age greater than 0.5 Gyr, the planet's interior would include 40-60 Earth masses of heavy elements according to standard planetary models (*23*). This would imply a composition of approximately half gas (hydrogen and helium) and half heavy elements (presumably ice and rock). Saturn, by contrast, is at least two-thirds gas by mass (*24*).

To investigate the long-term (secular) changes in the orbital parameters, and check on the system's stability, we integrated the best-fitting model forward in time by two million years. Within the context of our gravitational three-body model, secular variations occur on a timescale of about 40 years, without any significant excursions in orbital distance that would have led to instability. The planet's orbital eccentricity reaches a maximum of about 0.09. Likewise, the planet's line-of-sight orbital inclination changes by 0.2°, which is large enough that transits are only visible from Earth about 40% of the time (averaged over centuries). In particular, the planetary transits across star A should cease in early 2018, and return some time around 2042. The planetary transits across star B are already grazing, and are predicted to disappear for 35 years beginning in May 2014.

The planet experiences swings in insolation due to the motion of the stars on short timescales, and due to secular changes in the planet's orbit on long timescales. These variations are likely to affect the temperature and structure of the planet's atmosphere. The planet's current equilibrium temperature, averaged over several orbits, is between 170 and 200 K, assuming isotropic re-radiation of the stellar flux and a Bond albedo between 0.2-0.5 (in the neighborhood of Saturn's value of 0.34). Orbital motion of the stars and the planet are expected to produce seasonal temperature variations of around 30 K.

The planetary orbit is aligned with the stellar orbit to within 0.4°. This extreme coplanarity suggests that the planet was formed along with the stars, within a circumbinary protoplanetary disk, as opposed to being captured from another system. Planetesimal formation around an eccentric binary is a theoretical challenge, because of the large collision velocities of particles that are stirred by the stellar binary (*25*), although the detection of debris disks around close binaries has been interpreted as dust produced by colliding planetesimals (*26*). Subsequent stages of planet formation around binaries has been studied theoretically, both for terrestrial planets (*27*) and gas giants (*28*), but these and other theoretical studies (*29*) have lacked a well-specified circumbinary planetary system that could allow such a refinement of models.

Finally, the stars themselves are worthy of attention, independently of the planet. It is rare to measure the masses and radii of such small stars with such high precision, using geometrical and dynamical methods independent of stellar evolutionary models. In particular, Star B, with only 20% the mass of the Sun, is the smallest main-sequence star for which such precise mass and radius data are available (*30*). The mass ratio of 0.29 is also among the smallest known for binaries involving fully convective stars at the low-mass end of the main sequence (*29*). With well-characterized low-mass stars, in addition to a transiting circumbinary planet, this makes *Kepler*-16 a treasure for both exoplanetary and stellar astrophysical investigations.

42. NASA's Science Mission Directorate provided funding for the *Kepler* Discovery mission. LRD acknowledges the NASA *Kepler* Participating Scientist Program (grant number NNX08AR15G) and helpful discussions with the *Kepler* Science Team. JAC and DCF acknowledge support for this work was provided by NASA through Hubble Fellowship grants HF-51267.01-A and HF-51272.01-A awarded by the Space Telescope Science Institute, which is operated by the Association of Universities for Research in Astronomy, Inc., for NASA, under contract NAS 5-26555. JNW is grateful for support from the NASA Origins program (NNX09AB33G). GF acknowledges the support of the Hungarian OTKA grant MB08C 81013. The Kepler data used in this analysis can be downloaded from http://archive.stsci.edu/prepds/kepler_hlsp.

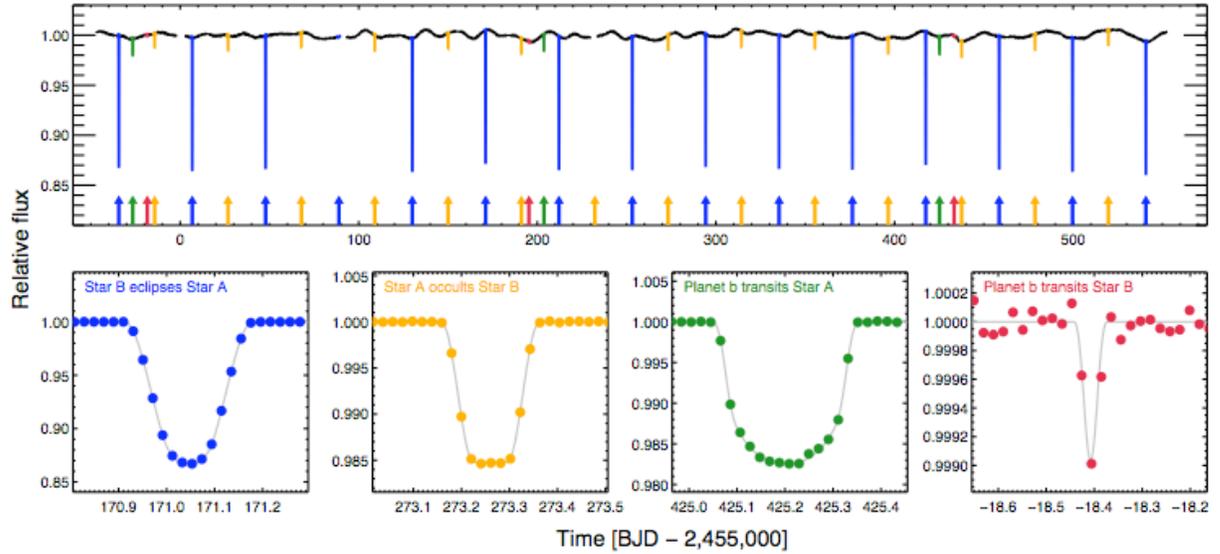

Figure 1. **Photometry of Kepler-16**.

*Top.*—Photometric time series from the *Kepler* spacecraft of star system *Kepler*-16 (KIC 12644769, KOI-1611, 2MASS 19161817+5145267, *Kepler* magnitude = 11.762). Each data point is the relative brightness at a given time (in barycentric Julian days, BJD). The 1% variations on ~10-day timescales are likely due to starspots carried around by stellar rotation (a periodogram gives a rotation period of about 35 days). The sharp dips are eclipses, appearing as vertical lines in this 600-day plot. They are identified as primary (B-eclipses-A; blue), secondary (A-occults-B; brown), tertiary (b-transits-A; green) and quaternary (b-transits-B; red). Because of interruptions in *Kepler* observing, data are missing from one primary eclipse at BJD 2,455,089, and one secondary eclipse at BJD 2,455,232. Note in particular the shifting order of the tertiary (green) and quaternary (red) eclipses: the first and third pairs begin with the tertiary eclipse, while the second pair leads with the quaternary eclipse. This is because the stars' orbital motion places them in different positions at each inferior conjunction of the planet. The stars silhouette the planet as they move behind it.

*Bottom.*—Close-ups (narrower scales in time and relative flux) of representative examples of each type of eclipse, along with the best-fitting model (gray), with parameters from Table 1.

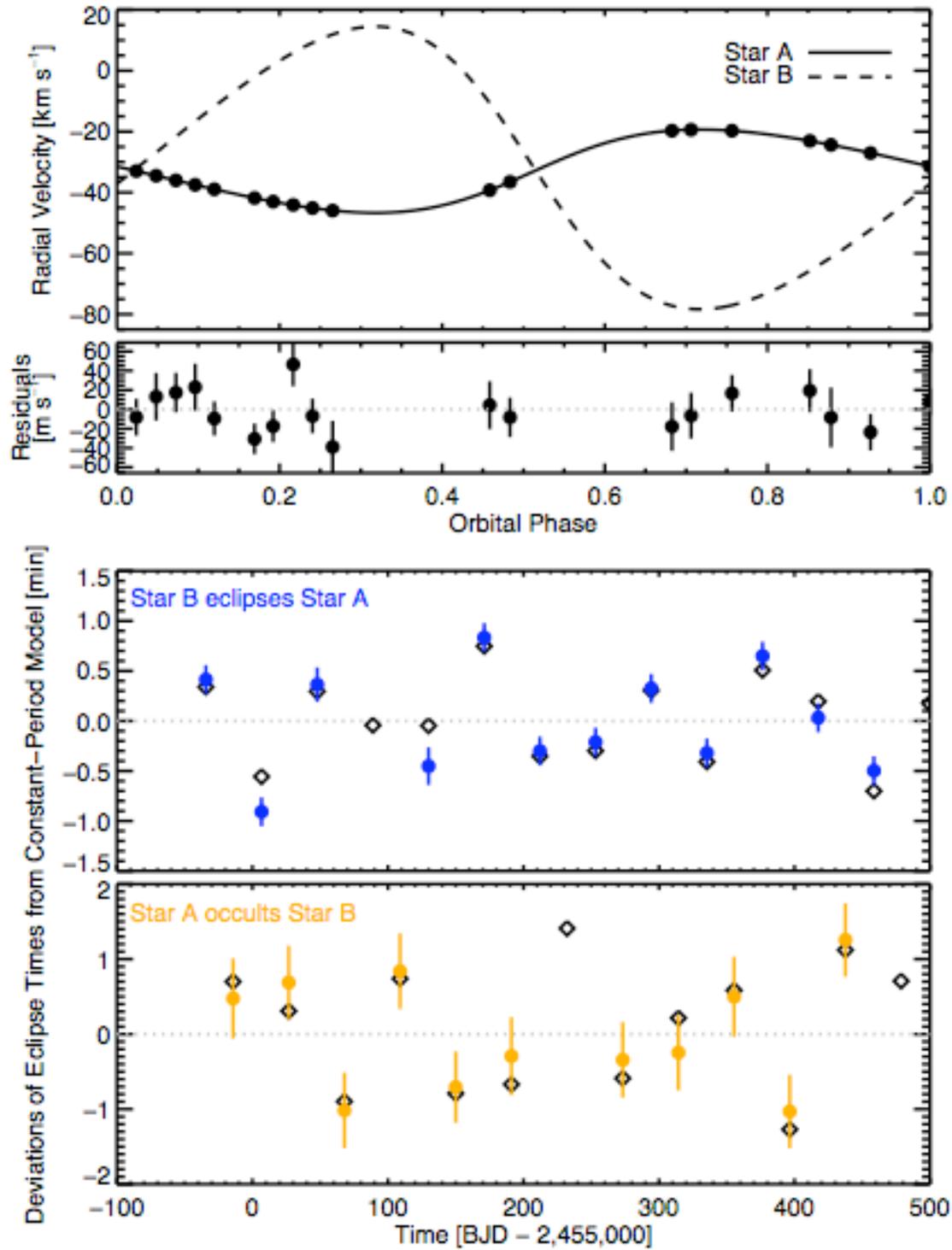

Figure 2. **Radial-velocity variations, and perturbations of eclipse times**.

*Top*.—Observed radial-velocity variations of star A as a function of orbital phase, based on observations with the TRES spectrograph and the Tillinghast 1.5m telescope at the Fred

Lawrence Whipple Observatory on Mt. Hopkins, Arizona (SOM). Solid dots are the data, and the smooth curve is the best-fitting model. Although only the light from star A could be detected in the spectra, the model for star B's motion is also shown. Residuals from the best model fits are given just below the radial velocity curve.

*Middle* and *Bottom panels.*—Deviations of the stellar eclipse times from strict periodicity, as observed (colored dots) and modeled (open diamonds). As noted previously, one primary eclipse and one secondary eclipse were missed. The deviations are on the order of one minute for both primary and secondary stellar eclipses. In the model, the effects of dynamical perturbations are dominant, with light-time variations contributing only at the level of one second. If the third body were more massive than a planet (> 13 Jovian masses), the timing variations would have exceeded 30 minutes. This would have been off the scale of the diagram shown here, and in contradiction with the observations.

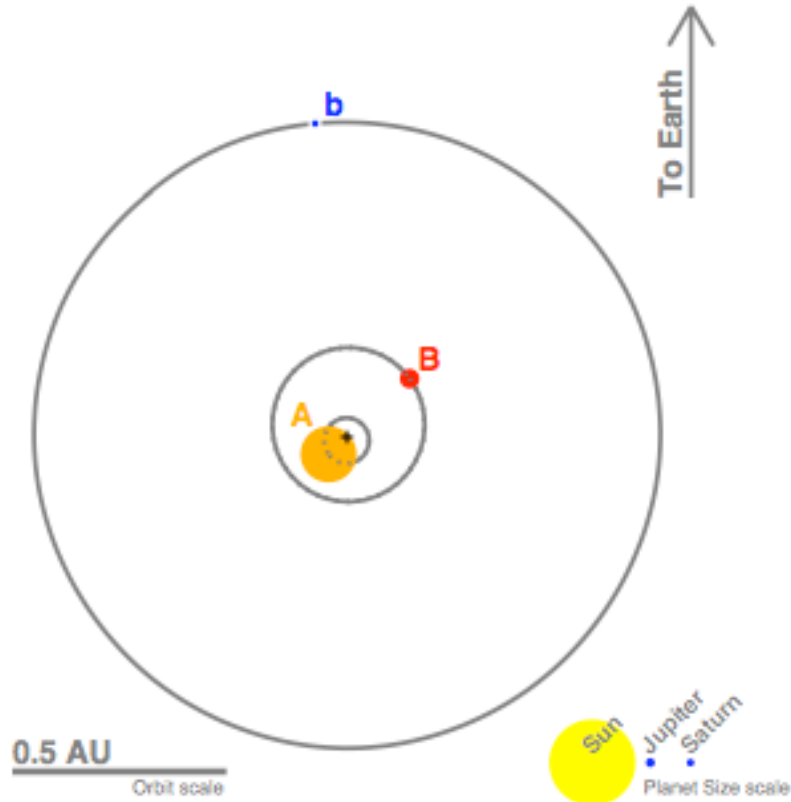

Figure 3: **Scale diagram of the *Kepler*-16 system**. The current orbits of the *Kepler*-16 system are shown as gray curves. The sizes of the bodies (including the Sun, Jupiter and Saturn) are in the correct proportions to one another, but they are on a scale 20 times larger than the orbital distance scale. We note that the binary and circumbinary planet orbital planes lie within 0.4° degree of each other (Table 1) so the orbits are essentially flat, as drawn. The planet's orbital eccentricity is nearly zero, while the orbital eccentricity of the binary star system is presently about 0.16. A "+" symbol marks the center of mass of all three bodies.

| Parameter | Value and Uncertainty |
|---|---|
| *Star A* | |
| Mass, $M_A$ ($M_\odot$) | $0.6897^{+0.0035}_{-0.0034}$ |
| Radius, $R_A$ ($R_\odot$) | $0.6489^{+0.0013}_{-0.0013}$ |
| Mean Density, $\rho_A$ (g cm$^{-3}$) | $3.563^{+0.017}_{-0.016}$ |
| Surface Gravity, $\log g_A$ (cgs) | $4.6527^{+0.0017}_{-0.0016}$ |
| Effective Temperature, $T_{\text{eff}}$ (K) | $4450 \pm 150$ |
| Metallicity, [m/H] | $-0.3 \pm 0.2$ |
| *Star B* | |
| Mass, $M_B$ ($M_\odot$) | $0.20255^{+0.00066}_{-0.00065}$ |
| Radius, $R_B$ ($R_\odot$) | $0.22623^{+0.00059}_{-0.00053}$ |
| Mean Density, $\rho_B$ (g cm$^{-3}$) | $24.69^{+0.13}_{-0.15}$ |
| Surface Gravity, $\log g_B$ (cgs) | $5.0358^{+0.0014}_{-0.0017}$ |
| *Planet b* | |
| Mass, $M_b$ ($M_{\text{Jupiter}}$) | $0.333^{+0.016}_{-0.016}$ |
| Radius, $R_b$ ($R_{\text{Jupiter}}$) | $0.7538^{+0.0025}_{-0.0023}$ |
| Mean Density, $\rho_b$ (g cm$^{-3}$) | $0.964^{+0.047}_{-0.046}$ |
| Surface Gravity, $g_b$ (m s$^{-2}$) | $14.52^{+0.70}_{-0.69}$ |
| *Binary star orbit* | |
| Period, $P_1$ (day) | $41.079220^{+0.000078}_{-0.000077}$ |
| Semi-major axis length, $a_1$ (AU) | $0.22431^{+0.00035}_{-0.00034}$ |
| Eccentricity, $e_1$ | $0.15944^{+0.00061}_{-0.00062}$ |
| Argument of Periapse, $\omega_1$ (deg) | $263.464^{+0.026}_{-0.027}$ |
| Mean Longitude, $\lambda_1$ (deg) | $92.3520^{+0.0011}_{-0.0011}$ |
| Inclination, $i_1$ (deg) | $90.3401^{+0.0016}_{-0.0019}$ |
| Longitude of Nodes, $\Omega_1$ (deg) | $\equiv 0$ (by definition) |
| *Circumbinary planet orbit* | |
| Period, $P_2$ (day) | $228.776^{+0.020}_{-0.037}$ |
| Semi-major axis length, $a_2$ (AU) | $0.7048^{+0.0011}_{-0.0011}$ |
| Eccentricity, $e_2$ | $0.0069^{+0.0010}_{-0.0015}$ |
| Argument of Periapse, $\omega_2$ (deg) | $318.^{+10.}_{-22.}$ |
| Mean Longitude, $\lambda_2$ (deg) | $106.51^{+0.32}_{-0.16}$ |
| Inclination, $i_2$ (deg) | $90.0322^{+0.0022}_{-0.0023}$ |
| Longitude of Nodes, $\Omega_2$ (deg) | $0.003^{-0.013}_{-0.013}$ |
| *Other parameters* | |
| Flux ratio of stars in *Kepler* bandpass, $F_B/F_A$ | $0.01555^{+0.00010}_{-0.00006}$ |
| Upper limit on third light (95% conf.), $F_X/(F_A+F_B)$ | $< 0.013$ |
| Radial Velocity Parameter of Barycenter, $\gamma$ (km s$^{-1}$) | $-32.769^{+0.035}_{-0.035}$ |
| Photometric Noise Parameter, $\sigma_{\text{phot}}$ | $0.0002403^{+0.0000062}_{-0.0000060}$ |

Table 1: **Characteristics of the Kepler-16 system.** For all parameters except $T_{\text{eff}}$ and [m/H] (which are described in the SOM) the results are based on the photometric-dynamical model. The quoted values and uncertainty intervals are based on the 15.85%, 50%, and 84.15% levels of the cumulative distributions of the marginalized posteriors for each parameter, making them analogous to "one-sigma" intervals for Gaussian statistics. The quoted orbital parameters are osculating Jacobian parameters at BJD 2,455,212.12316. The distance from *Kepler*-16 to Earth has not been measured, but is probably about 200 light years, judging from the apparent brightness of star A and theoretical models of stellar structure that give a crude estimate of its intrinsic luminosity.